\documentclass[aps,prb,twocolumn,showpacs,groupedaddress]{revtex4}
\usepackage{graphicx}

\begin{document}

\title{Quantum criticality in the phase diagram of K$_x$Sr$_{1-x}$Fe$_2$As$_2$: Evidence from transport and thermoelectric measurements}
\author{Melissa Gooch$^1$, Bing Lv$^2$, Bernd Lorenz$^1$, Arnold M. Guloy$^2$, and Ching-Wu Chu$^{1,3,4}$}
\affiliation{$^{1}$TCSUH and Department of Physics, University of Houston, Houston, TX 77204, USA} \affiliation{$^{2}$TCSUH and Department of
Chemistry, University of Houston, Houston, TX 77204, USA} \affiliation{$^{3}$Lawrence Berkeley National Laboratory, 1 Cyclotron Road, Berkeley,
CA 94720, USA} \affiliation{$^{4}$Hong Kong University of Science and Technology, Hong Kong, China}
\date{\today }

\begin{abstract}
% insert abstract here
The electrical transport and thermoelectric properties of K$_x$Sr$_{1-x}$Fe$_2$As$_2$ are investigated for 0$\leq$x$\leq$1. The resistivity
$\rho(T)$ shows a crossover from Fermi liquid-like temperature dependence at small x to linear $\varrho\sim T$ dependence at x$_c\simeq$0.4.
With further increasing x, $\varrho(T)$ becomes non-linear again. The thermoelectric power S(T) exhibits a similar crossover with increasing x
with a logarithmic T-dependence, S/T$\sim$ln(T), near the critical doping x$_c$. These results provide evidence for a quantum critical behavior
due to the coupling of low-energy conduction electrons to two-dimensional spin fluctuations.
\end{abstract}

 \maketitle

The discovery of superconductivity in quaternary rare-earth transition-metal oxypnictides with transition temperatures up to 55 K and the
existence of a spin density wave (SDW) phase in the non-superconducting parent
compounds\cite{kamihara:08,takahashi:08,chen:08b,chen:08,ren:08,ren:08b} has stimulated extensive discussions about the coexistence of both
states of matter, the role of magnetic order, its competition and mutual interaction with superconductivity, and the possible existence of
unusual normal state properties and a quantum critical point in the phase diagram of FeAs-based
compounds.\cite{giovannetti:08,xu:08,dai:08,ning:08,chu:08,margadonna:08,liu:08c,hess:08,kohama:08} The oxygen-free "122" compounds,
A$_x$Ae$_{1-x}$Fe$_2$As$_2$ with A=alkali metal and Ae=alkaline earth element, have been shown recently to form a solid solution over the whole
compositional range 0$\leq$x$\leq$1.\cite{sasmal:08,chen:08d,rotter:08b} The AeFe$_2$As$_2$ compounds are not superconducting but they exhibit a
SDW transition at 172 K, 140 K, and 205 K for Ae=Ca, Sr, and Ba, respectively.\cite{goldman:08,krellner:08,rotter:08c} With alkali metal doping
the SDW transition temperature decreases and superconductivity is observed above a critical concentration of about 0.17 in
K$_x$Sr$_{1-x}$Fe$_2$As$_2$\cite{sasmal:08} and $K_x$Ba$_{1-x}$Fe$_2$As$_2$.\cite{chen:08d,rotter:08b} With further increasing x (hole doping)
the superconducting T$_c$ passes through a maximum at 37 K and decreases to 3.7 K for x=1 (KFe$_2$As$_2$). The K$_x$Ae$_{1-x}$Fe$_2$As$_2$
superconductor is therefore the ideal system to probe the complete superconducting  phase diagram and the mutual correlation with the SDW order
and magnetic fluctuations.

It is not clear, however, if the magnetic and superconducting orders coexist and how the SDW phase boundary extends or extrapolates into the
superconducting phase with increasing doping. Careful investigations of the phase diagrams of K$_x$Sr$_{1-x}$Fe$_2$As$_2$\cite{gooch:08,lv:08}
and K$_x$Ba$_{1-x}$Fe$_2$As$_2$\cite{chen:08d,rotter:08b} suggest a narrow region between x$\simeq$0.17 and x$\simeq$0.25 where a SDW transition
is followed by a superconducting transition upon decreasing temperature. Although the magnetic order is apparently suppressed with the onset of
superconductivity, the extrapolation of the SDW phase boundary to T=0 would locate the quantum phase transition from the SDW phase to the
paramagnetic phase near a critical doping of x$_c\simeq$0.4. In order to follow the magnetic phase to T=0 and to investigate the possible
quantum critical point the superconductivity has to be suppressed, for example, by large external magnetic fields. However, this is difficult to
achieve since the critical field of the superconducting phase is very large.\cite{sasmal:08,lv:08} Alternatively, the physical properties can be
investigated in the temperature range above the suspected quantum critical point for the typical crossover signature that is expected in a
quantum critical regime.

We have therefore studied the normal state electrical and thermoelectric transport properties in the whole phase diagram of
K$_x$Sr$_{1-x}$Fe$_2$As$_2$ for 0$\leq$x$\leq$1. We observe a crossover from Fermi liquid behavior (x=0) to non-Fermi liquid temperature
dependence of resistivity and thermoelectric power close to the critical doping x$_c$. The results are consistent with the expected temperature
dependencies near a magnetic quantum critical point.

\begin{figure}[b]
\begin{center}
\includegraphics[angle=0, width=3 in]{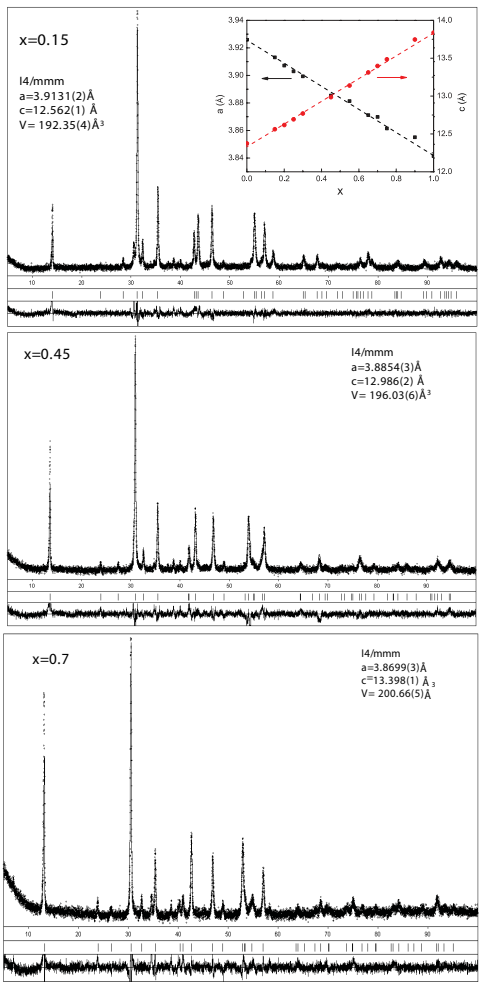}
\end{center}
\caption{(Color online) Selected X-ray spectra of the system K$_x$Sr$_{1-x}$Fe$_2$As$_2$ for x=0.15, 0.45, 0.7. The inset shows the change of
lattice parameters a and c with x.}
\end{figure}

Polycrystalline samples of K$_x$Sr$_{1-x}$Fe$_2$As$_2$ have been synthesized from the high-purity ternary compounds KFe$_2$As$_2$ and
SrFe$_2$As$_2$, as described earlier.\cite{sasmal:08} The careful mixing of the two end members at nominal ratios allows for a sensitive control
of the composition (doping). The X-ray analysis confirms the phase purity of the solid solutions, characteristic spectra and their corresponding
Rietveld refinement are shown in Fig. 1. All observed peaks are assigned to the I4/mmm structure. The continuous and nearly linear progression
of the lattice parameters with increasing x (c increasing and a decreasing, inset to Fig. 1) proves that a uniform solid solution forms for all
values of x. The resistivity was measured in a four probe configuration using the low-frequency (19 Hz) ac resistance bridge LR700 (Linear
Research). The thermoelectric power was measured employing a highly sensitive ac (0.1 Hz) method with a typical resolution of 0.05
$\mu$V/K.\cite{choi:01} The probe and the wires have been calibrated using a Hg high-T$_c$ superconductor below 130 K and high purity lead
between 130 and 300 K.

\begin{figure}
\begin{center}
\includegraphics[angle=0, width=2.5 in]{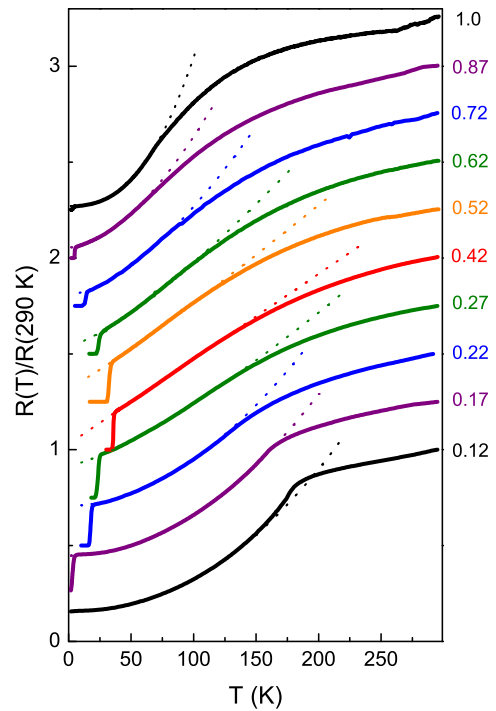}
\end{center}
\caption{(Color online) Resistivity of K$_x$Sr$_{1-x}$Fe$_2$As$_2$ vs. temperature for 0$\leq$x$\leq$1. For clarity different curves are
vertically offset by 0.25 units. The labels mark the values of x. The dotted lines show the fit to a power law, $\rho(T)=\rho_0+AT^n$ with the
exponents n shown in Fig. 3.}
\end{figure}

\begin{figure}
\begin{center}
\includegraphics[angle=0, width=2.5 in]{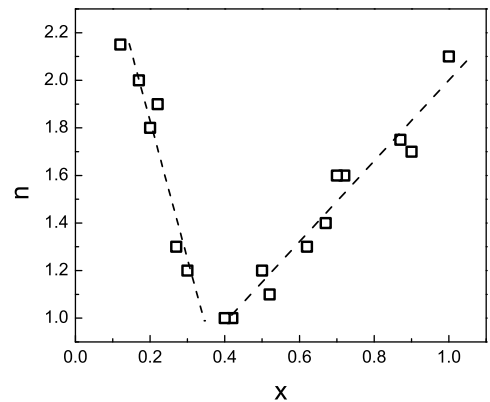}
\end{center}
\caption{The resistivity exponent n as a function of x.}
\end{figure}

The resistivities of the complete series of K$_x$Sr$_{1-x}$Fe$_2$As$_2$ are shown in Fig. 2. Different curves are offset by a constant unit of
0.25 for better clarity. Both ternary compounds, KFe$_2$As$_2$ and SrFe$_2$As$_2$, show a strongly non-linear temperature dependence of the
resistivity, $\rho$(T). However, approaching the critical doping of x$_c\simeq$0.4 the low-T part of the resistivity curve becomes increasingly
linear indicating a crossover into a non-Fermi liquid regime. Fitting the low-temperature part (above T$_c$) of $\rho$(T) to a simple power law,
$\rho(T)=\rho_0+AT^n$, an effective exponent $n$ can be extracted from the data of Fig. 2. Although the fitted exponent may have an error of the
order of 10 \% it provides a qualitative picture of the crossover of the transport properties. The power law dependence of $\rho$(T) with the
exponents n is shown by dotted lines in Fig. 2. It also defines the temperature range for which the power law is valid. The exponent $n$
decreases with x from about 2 to 1 near the critical x$_c$ and then increases again to values close to 2, as shown in Fig. 3. This crossover
behavior is consistent with the transport property expected at temperatures above a magnetic quantum critical point.\cite{vloehneysen:07} For
example, in the high-T$_c$ cuprate superconductors the linear $\rho$(T) was observed above the superconducting dome and the non-Fermi liquid
property was associated with the quantum critical point defined by the extrapolation of the pseudo gap temperature to T=0. In heavy Fermion
compounds, e.g. CeCu$_{6-x}$Au$_x$, the deviations from Fermi liquid properties have been observed frequently and the linear $\rho$(T) was
explained as a characteristic transport behavior in the quantum critical regime next to a magnetic quantum phase
transition.\cite{vloehneysen:98} The main difference between the heavy Fermion compounds and the K$_x$Sr$_{1-x}$Fe$_2$As$_2$ system is the
energy scale of the magnetic fluctuations. Whereas in CeCu$_{6-x}$Au$_x$ the typical magnetic transition temperatures are of the order of 1 K
the SDW transition in SrFe$_2$As$_2$ takes place at 200 K, two orders of magnitude higher. Therefore, the quantum critical region in which clear
deviations from the Fermi liquid model can be observed extends to higher temperatures.

The resistivity data presented in Fig. 2 and discussed above have been acquired for dense polycrystalline samples. It is not known whether the
grain boundary contributions could affect the T-dependence of $\rho$ and the exponent n in Fig. 3. The thermoelectric power S, however, is a
zero current property and does not depend on grain boundaries as much as the resistivity. The temperature dependent data of S(T) throughout the
phase diagram of K$_x$Sr$_{1-x}$Fe$_2$As$_2$ are summarized in Fig. 4. S(T)/T is shown as a function of ln(T). All data can be separated into
two groups, each group showing a consistent and characteristic T-dependence: (i) For x$<$0.3 the thermoelectric power S/T exhibits an almost
identical T-dependence above the SDW transition. This is obvious from the data shown in Fig. 4a. The dashed circle indicates the relevant
T-range above T$_{SDW}$. (ii) For x$>$0.5 the temperature dependence of S/T is similar for all x above the superconducting transition, as shown
in Fig. 4b. The dashed circle emphasizes on this temperature range and the closeness of all data in the high doping range. S(T)/T as a function
of ln(T) is non-linear in both cases, for x$<$0.3 and x$>$0.5. However, between the doping of 0.3 and 0.5 the thermoelectric power shows a
distinct crossover between the characteristic T-dependence of group 1 (Fig. 4a) and group 2 (Fig. 4b). For x$\simeq$0.4 S/T varies linearly with
ln(T) over a large temperature range, from the superconducting T$_c\simeq$36 K to about 180 K. The data in the crossover range are shown in more
detail in Fig. 4c. The dashed line attached to the x=0.42 data shows that S/T$\sim$ln(T) at this critical doping. The scaling property of the
thermoelectric power, as revealed in Fig. 4c, is a strong signature of quantum criticality and it is consistent with the non-Fermi liquid like
temperature dependence of the resistivity. It should be noted that a similar logarithmic dependence of S/T has been reported very recently in
the hole-doped high-T$_c$ cuprate superconductor, La$_{1.4-x}$Nd$_{0.4}$Sr$_x$CuO$_4$.\cite{daou:08}

\begin{figure}
\begin{center}
\includegraphics[angle=0, width=2.5 in]{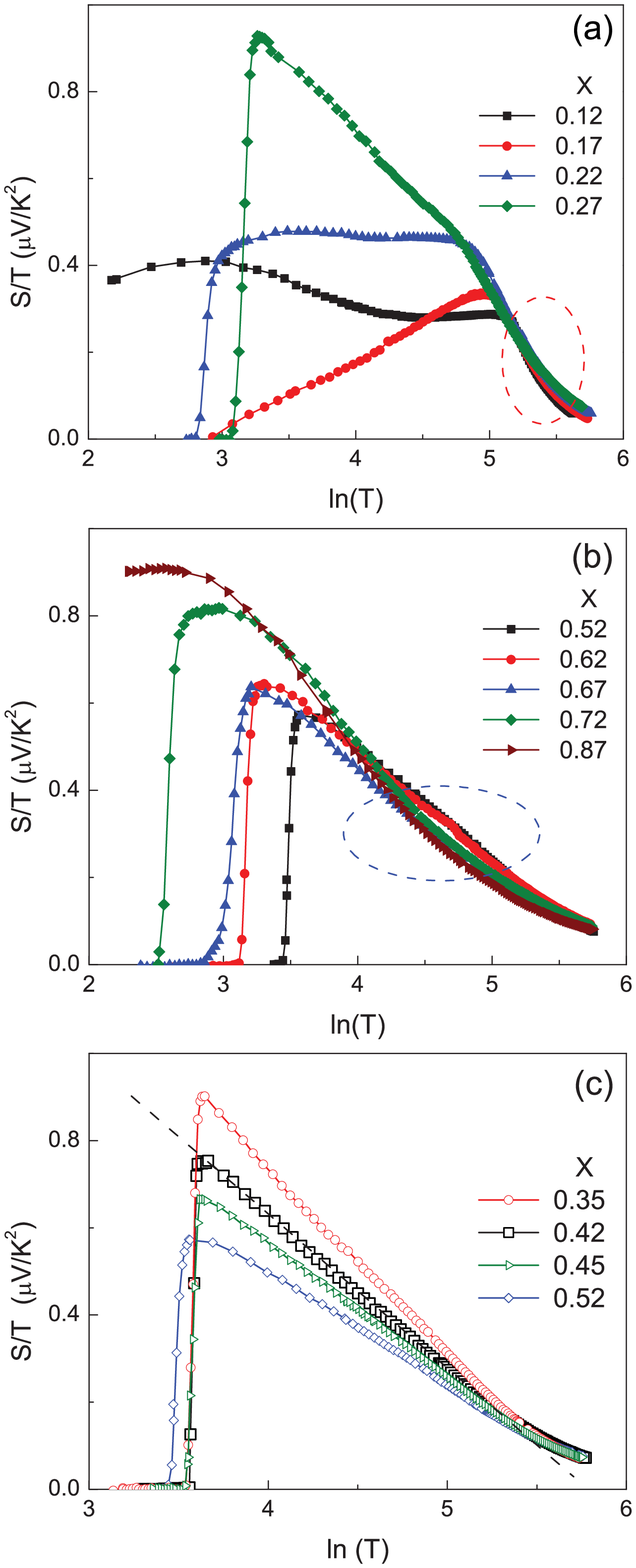}
\end{center}
\caption{(Color online) Scaling plot S/T vs. ln(T) of the thermoelectric power in the phase diagram of K$_x$Sr$_{1-x}$Fe$_2$As$_2$. (a) S/T for
x$<$0.3, (b) S/T for x$>$0.5, and (c) S/T in the crossover region between x=0.3 and x=0.5. The dashed line in (c) shows the logarithmic scaling
at the critical doping x$_c$.}
\end{figure}

Thermodynamic and transport properties near magnetic quantum phase transitions have been investigated mostly in heavy Fermion systems like the
above mentioned CeCu$_{6-x}$Au$_x$. Whereas the undoped parent compound, CeCu$_6$, is nonmagnetic, the substitution of Au induces incommensurate
antiferromagnetic order for x$>$0.1. Right at the critical concentration of x=0.1 the CeCu$_{6-x}$Au$_x$ system exhibits quantum critical
scaling at finite temperatures as expressed in a linear T-dependence of resistivity and a logarithmic temperature dependence of the heat
capacity, C/T.\cite{vloehneysen:98} Similar scaling properties have been found in other heavy Fermion compounds with a magnetic quantum critical
point.\cite{vloehneysen:07} Since the typical scaling range in heavy Fermion compounds is of the order of 1 K the heat capacity is the
entropy-related quantity that is experimentally accessible with the precision required to prove the quantum critical behavior (note that at
higher temperatures the lattice contribution to the heat capacity masks the electronic part and makes it far more difficult, if not impossible,
to extract accurately the electronic heat capacity that is expected to follow the scaling laws). In our high-T$_c$ K$_x$Sr$_{1-x}$Fe$_2$As$_2$
system, however, the low temperature region is not accessible because the superconducting state is stabilized and dominates the physical
properties below T$_c$. Therefore, other physical quantities, like resistivity or thermoelectric power, had to be investigated with respect to
quantum critical scaling properties.

The thermoelectric property near a quantum phase transition is less well investigated as compared to resistivity or heat capacity. For
CeCu$_{6-x}$Au$_x$ (x=0.1) at low temperatures S(T) was reported to vary nonlinearly with T in contrast to the linear dependence expected for a
Fermi liquid.\cite{benz:99} Theoretically, logarithmic scaling in the quantum critical regime of, for example, C/T is expected if the dimension
of the critical fluctuations (d) equals the dynamical critical exponent (z).\cite{vloehneysen:07} The critical scaling properties of
CeCu$_{5.9}$Au$_{0.1}$ have been explained based on a scaling theory for d=2 and z=2.\cite{rosch:97} Alternatively, based on a spin-Fermion
model proposed by Abanov and Chubukov,\cite{abanov:00} it has been shown that low-energy conduction electrons interacting with quasi-2d spin
fluctuations give rise to a linear with temperature resistivity, a logarithmic T-dependence of the heat capacity C/T, and a similar logarithmic
scaling of the thermoelectric power, S/T$\sim$ln(T).\cite{paul:01} The ln(T)-dependence of S/T for K$_{0.42}$Sr$_{0.58}$Fe$_2$As$_2$ shown in
Fig. 4b as well as the T-linear resistivity (Figs. 2 and 3) are consistent with the model of low-energy conduction electrons interacting with
quasi-2d spin fluctuations. The lower dimensionality of the magnetic fluctuations finds its natural origin in the layered structure of the FeAs
compounds with the magnetic Fe-ions confined to the Fe$_2$As$_2$ layers.\cite{qi:08}

One question remaining is the origin of the magnetism and free carriers in FeAs compounds. While in typical heavy Fermion systems (e.g.
CeCu$_6$) magnetic moments are introduced through localized f-electrons of the rare earth ions and conduction electrons are provided by
transition metals, this is not necessarily the case in FeAs compounds since the electrons at the Fermi surface are mainly from hybridized
orbitals of Fe and As with mainly d- or p-electron character. In the present K$_x$Sr$_{1-x}$Fe$_2$As$_2$ system there is no alternative source
for magnetic moments than the Fe ions. A recent study of the "undoped" parent compounds proposed a separation of the electronic excitations into
an "incoherent" part, further away from the Fermi surface and giving rise to local magnetic moments interacting with each other through
frustrated superexchange coupling, and a "coherent" part in the vicinity of the Fermi surface.\cite{dai:08} Recent electron spin resonance
experiments on La(O,F)FeAs seem to support the existence of local magnetic moments, their coupling to itinerant electrons, and the presence of
strong magnetic frustration.\cite{wu:08d} The coherent carriers couple to the local moments and compete with the SDW order. Increasing the
carrier concentration by doping can tune the system to a magnetic quantum phase transition. The theoretical treatment within a low energy
Ginzburg-Landau theory indeed describes an antiferromagnetic quantum phase transition with the dynamical critical exponent z=2 and the effective
dimension d+z=4.\cite{dai:08} However, the dimension and the nature of the spin fluctuations are important since logarithmic scaling is expected
only for d=2, z=2 or d=3, z=3.\cite{vloehneysen:07} For three dimensional magnetic fluctuations (d=3) the dynamical critical exponent has to be
z=3 to explain the logarithmic scaling, as for example in the case of a metallic ferromagnet. The existence of ferromagnetic spin fluctuations
in LaO$_{1-x}$F$_x$FeAs has recently been proposed based on the large Wilson ratio.\cite{kohama:08} Which scenario applies to the system
K$_x$Sr$_{1-x}$Fe$_2$As$_2$ remains an open question and has to be resolved in future investigations.

Our results obtained for the K$_x$Sr$_{1-x}$Fe$_2$As$_2$ system, linear T-dependence of resistivity and logarithmic T-dependence of S/T, are
consistent with the expected scaling behavior of various thermodynamic and transport properties in the critical regime near a magnetic quantum
phase transition.\cite{vloehneysen:07} The crossover properties of $\rho$(T) and S/T throughout the phase diagram of K$_x$Sr$_{1-x}$Fe$_2$As$_2$
and the scaling behavior near the critical doping x$_c$ provide strong experimental support for the existence of a magnetic quantum critical
regime above the superconducting T$_c$. This regime needs to be investigated further by extending the accessible temperature range to lower T.
To this end the superconducting phase has to be suppressed. Since the critical fields appear to be too high to suppress the superconducting
state, other control parameters such as pressure or chemical substitution may be considered as an alternative.

\begin{acknowledgments}
Stimulating discussions with S. Wirth and Q. Si are gratefully acknowledged. This work is supported in part by the T.L.L. Temple Foundation, the
J.J. and R. Moores Endowment, the State of Texas through TCSUH, the USAF Office of Scientific Research, and at LBNL through USDOE. A.M.G. and
B.L. acknowledge the support from the NSF (CHE-0616805) and the R.A. Welch Foundation.
\end{acknowledgments}

\bibliographystyle{phpf}

\end{document}